\documentclass[a4paper,12pt]{article}
\usepackage[margin=2cm]{geometry}
\usepackage[utf8]{inputenc}
\usepackage[T2A]{fontenc}
\usepackage{amsmath}
\usepackage{amssymb}
\usepackage{bm}
\usepackage[dvipdf]{graphicx}
\usepackage{authblk}
\usepackage{cite}
\usepackage{indentfirst}

\newcommand{\vecb}[1]{{\bm{\mathrm{#1}}}}
\newcommand{\Dfrac}[2]{\frac{d#1}{d#2}}

\newcommand{\DPfrac}[2]{\frac{\partial#1}{\partial#2}}

\begin{document}

\title{On the possible turbulence mechanism in accretion disks in non-magnetic
  binary stars}

\author[1]{E.~P.~Kurbatov\thanks{kurbatov@inasan.ru}}
\author[1]{D.~V.~Bisikalo\thanks{bisikalo@inasan.ru}}
\author[1]{P.~V.~Kaygorodov\thanks{pasha@inasan.ru}}
\affil[1]{Institute of Astronomy, Russian Acad. Sci.}

\maketitle

\begin{abstract}
The arising of turbulence in gas-dynamic (non-magnetic) accretion disks is a
major issue of modern astrophysics. Such accretion disks should be stable
against the turbulence generation, in contradiction to observations. Searching
for possible instabilities leading to the turbulization of gas-dynamic disks is
one of the challenging astrophysical problems. In 2004, we showed that in
accretion disks in binary stars the so-called precessional density wave forms
and induces additional density and velocity gradients in the disk. Linear
analysis of the fluid instability of an accretion disk in a binary system
revealed that the presence of the precessional wave in the disk due to tidal
interaction with the binary companion gives rise to instability of radial modes
with the characteristic growth time of tenths and hundredths of the binary
orbital period. The radial velocity gradient in the precessional wave is shown
to be responsible for the instability.  A perturbation becomes unstable if the
velocity variation the perturbation wavelength scale is about or higher than
the sound speed. Unstable perturbations arise in the inner part of the disk
and, by propagating towards its outer edge, lead to the disk turbulence with
parameters corresponding to observations (the Shakura-Sunyaev parameter $\alpha
\lesssim 0.01$).

\end{abstract}

\section{Introduction}

High accretion rates observed in accretion disks in binary stars can be
explained only by the presence of turbulent viscosity
\cite{Shakura1972AZh....49..921S,Shakura1973A&A....24..337S,%
Lynden-Bell1974MNRAS.168..603L}
(see also a historical review~\cite{Shakura2014A&A....57..407S}).
The turbulence itself should arise due to some instability
\cite{Shakura1972AZh....49..921S,Lynden-Bell1974MNRAS.168..603L}. For a long
time, attempts have been made to search for fluid instabilities in Keplerian
disks (see, for example, references in paper
\cite{Godon1999ApJ...521..319G}). However, it can be shown that in a Keplerian
disk small radial perturbations are stable according to the Rayleigh
criterion~\cite{Balbus1996ApJ...467...76B}. In addition, numerical
simulations~\cite{Balbus1996ApJ...467...76B} suggest that azimuthal short
wavelength modes do not display instability as well. The numerical study of
long wavelength perturbations in a thin Keplerian
disk~\cite{Godon1999ApJ...521..319G} revealed that such perturbations grow to
become non-linear and then decay not quenching the disk turbulence.

Many authors have applied the magneto-rotational instability (MRI)
\cite{Velikhov1959ZETF,Chandrasekhar1960PNAS...46..253C} to accretion
disks~\cite{Balbus1991ApJ...376..214B}. However, this type of instability as
the reason for disk turbulence meets some difficulties: (i) in the majority of
close binary stars there is no observational evidence for the magnetic field;
(ii)  the disk turbulence due to MRI requires the presence of a seed magnetic
field; (iii) the magnetic field growth stabilizes perturbations, i.e. suppress the
instability~\cite{Chandrasekhar1960PNAS...46..253C}; (iv) at the non-linear
stage MRI saturates and, as a consequence, the angular momentum transfer
through the disk significantly decreases~\cite{Regev2009EAS....38..165R}. In
addition, recently the very existence of MRI in thin disks was
questioned~\cite{Liverts2012PhRvL.109v4501L}.

There have been several papers that further examined the fluid turbulence in
the disks. For example, in paper~\cite{Fridman2008UFN} the turbulence was
proposed to arise due to the super-reflection
instability. Papers~\cite{Cabot1984ApJ...277..806C,Klahr2003ApJ...582..869K}
argued that disks with negative entropy gradient (in the presence of radiative
cooling) can be subjected to baroclinic instability for axially non-symmetric
modes. Among the recent papers on gas-dynamic turbulence in astrophysical disks
we can notice paper~\cite{Mukhopadhyay2013JPhA...46c5501M}, which used the
statistical approach to the turbulence modeling. In such models, the field of
pressure fluctuations is represented by a stochastic force in equations of
motion. Due to simplicity and generality, such a source is usually modeled as a
Gaussian random process that is delta-correlated in time. In our opinion, this
approach may help in studying only well developed turbulence, but not the
process of its arising, as claimed by the
authors~\cite{Mukhopadhyay2013JPhA...46c5501M}. The reason is that this random
force is not consistent with gas-dynamic equations, therefore the growth of
perturbations due to this force cannot be viewed as the appearance of fluid
instability.

The study of instabilities is usually carried out in the frame of certain a
priori assumptions on the structure and parameters of the accretion disk. In
particular, one usually assumes a near-Keplerian velocity distribution in the
disk, its homogeneity in the equatorial plane, as well as circular form of the
stream lines or their weak eccentricity. In fact, such idealized assumptions
can be invalid already within the ‘pure’ gas-dynamic frame. For example, in
paper~\cite{Lyubarskij1994MNRAS.266..583L} it was shown that in a viscous
accretion disk, orbits of particles are unstable with respect to the
eccentricity growth, so that the disk ellipticity increases. In accretion disks
in binary stars specific physical conditions can occur, including shocks, tidal
interaction, resonances. These features can significantly affect the gas flow,
instability growth, turbulence and angular momentum transfer. In accretion
disks in semi-detached binaries, steady shocks arise due to the tidal
interaction with the secondary
component~\cite{Sawada-et-al:219:86,Sawada-et-al:221:86,Sawada-et-al:87} and
the interaction of the gas stream from the inner Lagrangian point with the
circumdisk halo~\cite{Bisikalo1997ARep...41..786B,Bisikalo1997ARep...41..794B,%
Bisikalo1998ARep...42..621B,Bisikalo-et-al:98,Bisikalo1999ARep...43..229B,%
Bisikalo1999ARep...43..587B,Bisikalo1999ARep...43..797B,%
Bisikalo2000ARep...44...26B,Bisikalo-et-al:2000,Bisikalo2001ARep...45..611B,%
Bisikalo2001ARep...45..676B,Molteni-et-al:2001,BinaryStars:2002,%
Fridman2003PhLA..317..181F}.

Numerical simulations also reveled that the tidal interaction with the
secondary component leads to arising of a specific type of waves in accretion
disks --- the precessional density
waves~\cite{Bisikalo2004ARep...48..449B}. Waves of this kind have spiral form
and occur virtually in the entire disk. Qualitatively, such a wave can be
represented as the envelope of a family of elliptical orbits precessing in a
non-symmetric gravitational field~\cite{Bisikalo2004ARep...48..449B}. The
ellipticity of orbits can result from the eccentric instability which arises
due to either viscous forces ~\cite{Lyubarskij1994MNRAS.266..583L} or
resonances~\cite{Lubow1991ApJ...381..259L}. In the last case, linear modes are
excited in the disk, and the single-arm spiral with azimuthal wave number
$m = 1$ has the maximum increment~\cite{Lubow1991ApJ...381..259L}.

Stability of the disk in which the model $m = 1$, is present has been studied
earlier. For example, paper~\cite{Kato2008PASJ...60..111K} considered the
interaction of linear perturbations with originally specified linear mode and
in the approximation of large radial wave numbers. In that paper it was shown
that perturbations propagating within the disk plane can be unstable, and the
increment, according to the authors’ opinion, can exceed the Keplerian
frequency in the disk.

The present paper is devoted to the study of instability of radial (axially
symmetric) perturbations, which form in the presence of the precessional
density wave, and the wave itself here is given as a numerical solution of
gas-dynamic equations. That the wave is tightly wounded allows us to neglect
the angular dependence of all considered values for each radial direction
chosen. As a result, we shall be restricted by analysis of the radial
perturbation modes only. An important feature of the precessional wave is that
it represents a smooth solution, which provides fast convergence of spectral
methods used in the stability analysis. The presence of gradients at the wave
significantly changes the dispersion relation for linear perturbations and
leads to conditions facilitating the instability growth.

In Section 2 a scenario of the density wave arising is presented. In Section 3
the linear analysis of perturbations in a thin isothermal Keplerian disk is
introduced and applied to the numerical model of the accretion disk with
precessional wave. In Section 4, a physical analysis of the results is
performed. The discussion of results and conclusion are given in Section 5.

\section{Precessional density wave}
\label{sec:prec_wave}

Accretion disks in semi-detached binaries have a sufficiently complicated
structure, since the tidal forces and the disk interaction with the
inter-component envelope and the gas stream from the inner Lagrangian point
$L_1$ deform the disk shape and lead to the formation of different
shocks. Tidal shocks and the ‘hot line’ --- the region of impact of the gas
stream from the inner Lgrangian point $L_1$ with the circumdisk halo –-- mostly
affect the flow. Three-dimensional hydrodynamic simulations demonstrate that
these waves do not penetrate deep inside the cold (with a temperature of $\sim
10^4$~K) accretion disk and leave most of the disk weakly perturbed, which
creates conditions for the third-type wave, the precessional density wave, to
appear~\cite{Bisikalo2004ARep...48..449B}. The form of stream lines in the
accretion disk is close to the corresponding elliptical Keplerian orbits with
the accreting star residing in one of the focuses. This can be explained by the
fact that in the region free from strong gas-dynamic perturbations due to
steady shocks, the disk is almost homogeneous and gravitational forces dominate
over forces due to gas pressure gradients in the equatorial plane of the
system.  The tidal interaction increases the stream line eccentricity in the
disk and forces their apse line to counter rotate the disk. The tidal forces
act on the stream lines non-homogeneously: the outer stream lines tend to
rotate faster than the inner ones, since in a gas disk (at least if the Knudsen
number%
\footnote{%
The Knudsen number here is the ratio of the particle free-path length to the
characteristic scale of the problem, which here can be the disk thickness. For
the typical number density in the accretion disk of the order of
$10^{11}$~cm$^{-3}$ the free path length is $10^{-10}$~a.u., while the disk
thickness is $\lesssim 10^{-2}$~a.u.
}
is much less than one) there can be no intersecting stream lines, some mean
precessional velocity is established, which is one and the same for all stream
lines. This velocity can be approximately estimated using the
formula~\cite{Kumar:86,Warner:95}:
\begin{equation}
  \dfrac{P_\mathrm{pr}}{P_\mathrm{orb}} \simeq
  \dfrac{4}{3} \dfrac{(1+q)^{1/2}}{q} \left( \dfrac{r}{A} \right)^{-3/2} \,,
  \label{eq:precession}
\end{equation}
where $P_\mathrm{pr}$ is the precessional period; $P_\mathrm{orb}$ is the
system's orbital period; $q$ is the binary components mass ratio; $r$ is the
characteristic size of the orbit; $A$ is the distance between the binary
components.

\begin{figure}
  \centering
  \includegraphics[width=8cm]{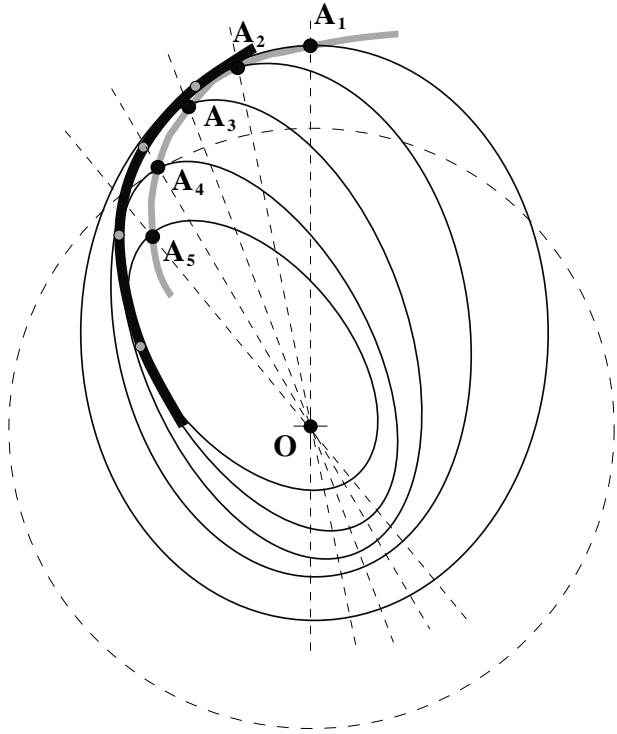}
  \caption{Schematics of the spiral pattern formation in the inner unperturbed
    parts of cold gas disks. The accretor is at the center (O), $A_1 \dots A_5$
    denote the stream line apastrons.}
  \label{fig:flowlines}
\end{figure}

The approaching of stream lines that moves with different velocities leads to
the formation of a spiral pattern shown schematically in
Fig.~\ref{fig:flowlines}.
The mater velocity along the stream line, whose form is primarily determined by
gravitation, will be close to the local Keplerian velocity for the
corresponding (elliptical) orbit. Accordingly, the velocity will be minimal at
the stream line apastron. However, as the flux value should be conserved along
the stream line, the matter density should also change along the stream line
and reach maximum at the apastron. Therefore, the spiral arm formed by the
stream lines apastrons will look like the density wave, as was shown in
paper~\cite{Bisikalo2004ARep...48..449B}.

In the observer’s reference frame the precessional wave is almost steady –--
its shifts by $1^\circ \-- 3^\circ$ in the retrograde direction in one orbital
period~\cite{Bisikalo2004ARep...48..449B}. Thus, the density and velocity
distribution in the wave can be considered stationary on time scales of the
order of several tens of the characteristic disk periods.

\begin{figure}
  \centering
  \includegraphics{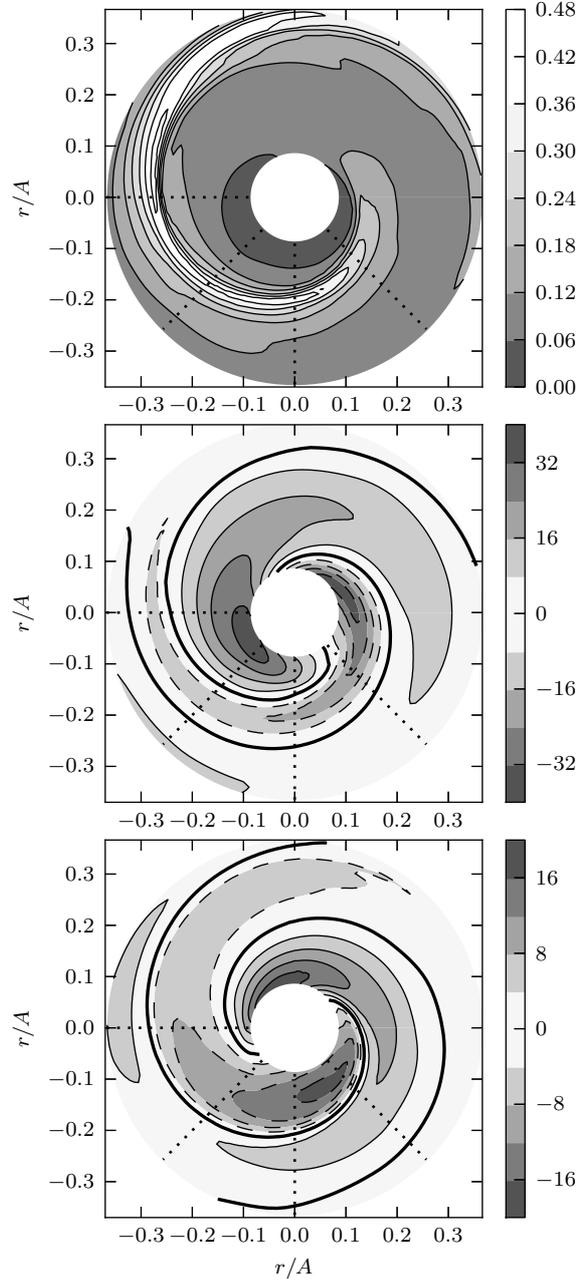}
  \caption{Maps of the  surface density (upper panel), radial velocity (middle
    panel) and deviations of the angular velocity from the Keplerian profile
    (bottom panel) according to the numerical
    model~\cite{Bisikalo2004ARep...48..449B}. Velocities are in units of the
    sound speed     $c_\mathrm{T}$. The thick lines show zero velocity levels.}
  \label{fig:disk_in}
\end{figure}

Fig.~\ref{fig:disk_in}
shows distributions of the surface density and the radial and angular velocity
in the numerical simulations of the disk in a close binary
system~\cite{Bisikalo2004ARep...48..449B}. The calculations were carried out
for the binary with parameters: accretor’s mass $M_1 = 1~M_\odot$, donor's mass
$M_2 = 0.05~M_\odot$, binary separation $A = 0.625~R_\odot$, binary orbital
perios $P_\mathrm{orb} = 4830$~s. The precessional wave is distinctly seen as a
spiral-like overdensity in the surface density map. In the radial velocity map
the region with the density wave is bounded by the zero radial velocity
lines. The zero tangential velocity lines coincide with radial velocity
extrema.

\begin{figure}
  \centering
  \includegraphics{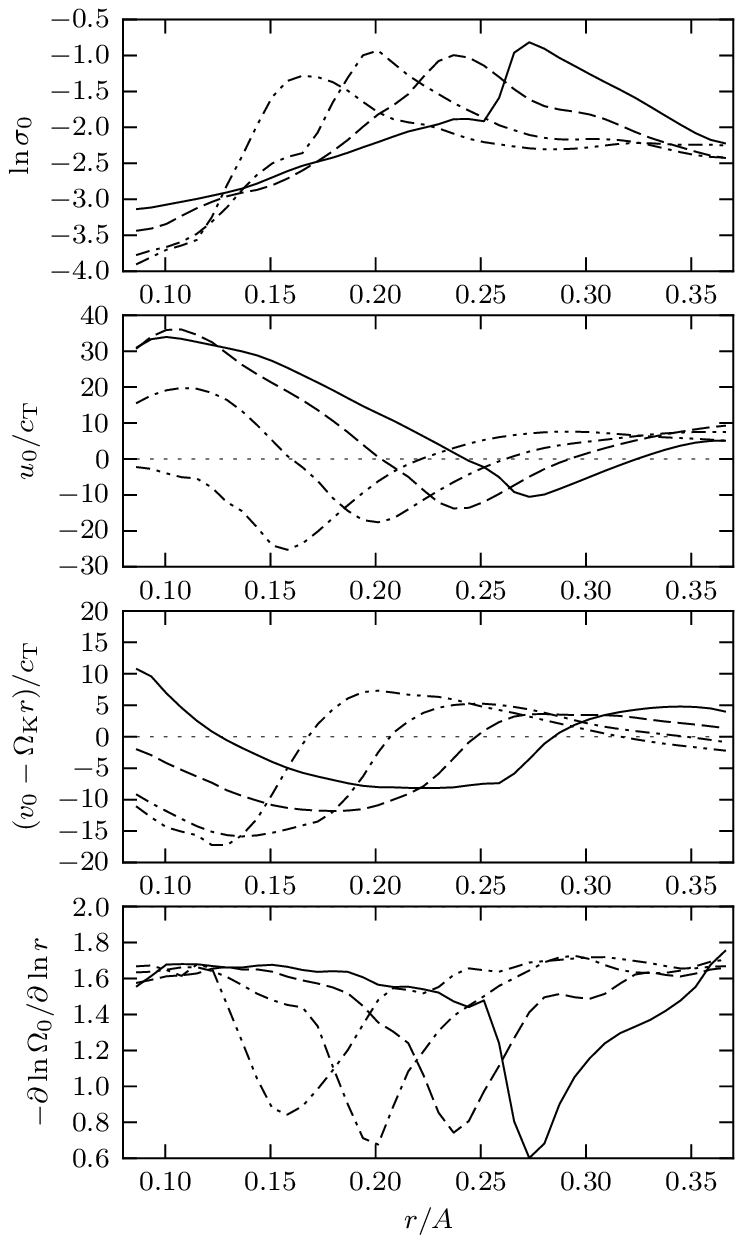}
  \caption{Plots of the surface density (a), radial velocity (b), angular
    velocity (c) and the rotational curve slope (d) in the numerical accretion
    disk  model~\cite{Bisikalo2004ARep...48..449B}. The plots correspond to the
    radial disk cuts in the directions $180^\circ$ (the solid curves),
    $225^\circ$ (the dashed curves), $270^\circ$ (the dash-dotted curves) and
    $315^\circ$ (the dash-double-dotted curves).}
  \label{fig:cuts_in}
\end{figure}

The surface density, radial and tangential velocity distributions are shown in
Fig.~\ref{fig:cuts_in}
along four radial directions in the disk: $180^\circ$, $225^\circ$,
$270^\circ$, and $315^\circ$ ($0^\circ$ corresponds to the direction from the
secondary component along the line connecting the binary components
centers). The density peaks observed at phases $0.15 \-- 0.25$ correspond to
intersections of the profiles with the precessional wave and well correlate
with the radial velocity minima. The radial and tangential velocity profiles
are significantly different along different directions. Nevertheless, these
profiles share common features determined by properties of elliptical
orbits. For example, for all profiles the tangential velocity is, on average,
sub-Keplerian (as accretion occurs through the disk), however in the outer
parts of the disk the tangential velocity is much smaller than in the inner
parts, since the velocity of matter moving along a Keplerian orbit decreases
with distance from the star. The radial velocity distribution significantly
depends on the direction –-- three profiles presented in
Fig.~\ref{fig:cuts_in}b demonstrate positive (i.e. directed away from the
accretor) velocity in the inner parts of the disk, at the same time the radial
velocity in the fourth profile is negative in the same region. This behavior is
clear, since the matter moving along an eccentric orbit can be closer to or
further away from the star at different parts of the trajectory. Nevertheless,
it should be noted that in the outer parts of the disk the radial velocity is
positive for all profiles, since the angular momentum of the disk decreases in
the outer parts due to decretion of matter. It should be specially stressed
that the radial velocity difference over the disk is fairly large and along
some directions can be as large as several tens of sound speed.

As follows from the distributions presented in Fig.~\ref{fig:disk_in}, the
precessional wave is tightly wounded. This allows one approximately treat
perturbations caused by the wave as axially symmetric. This approximation
cannot be valid in the central parts of the disk, where the axial symmetry is
significantly violated (it is the most clearly seen in the velocity
distribution). In what follows, we shall analyze perturbations excluding the
central part of the disk with radius smaller than $0.08~A$ from
calculations. The results of calculations suggest that the outer parts of the
disk are subjected to strong gas-dynamic perturbations.   At shocks located
close to the disk edge, the stream lines are broken, and correspondingly the
methods we use in the present paper cannot be applied to study instabilities in
these regions, since the calculation domain was limited by the radius
$0.37~A$.

\section{Linear analysis of perturbations in a disk with precessional wave}
\label{sec:analysis}

The density and radial and tangential velocity distributions obtained in the
numerical simulations~\cite{Bisikalo2004ARep...48..449B}, which were described in
Section~\ref{sec:prec_wave}, were taken as the background distributions for the
analysis of perturbations. Perturbations propagating in the direction
perpendicular to the disk are sound perturbations, i.e. they do not show
instability. Therefore, it seems plausible to assume that the account for
vertical perturbations can lead to the appearance of additional sound modes and
only insignificantly change frequencies and increments of radial
perturbations. These considerations allow significant simplifications of
calculations to be made by excluding the vertical degree of freedom. All the
subsequent analysis is carried out in two dimensions. Therefore, the results of
numerical simulations were reduced to two dimensions by integrating
distributions along the direction perpendicular to the
disk~\cite{Gor'kavyj1994pprc.book.....G}.

The calculations were performed in the inertial reference frame, where the
precessional wave is almost at rest. The presence of the time-dependent
gravitational field of the secondary component in this frame, generally
speaking, can give rise to additional spiral pattern co-moving with the
rotation~\cite{Kley2008A&A...487..671K}. The amplitude of deviations in the
density and radial and tangential velocity distributions due to this effect is
scaled with the components mass ratio~\cite{Kley2008A&A...487..671K}, which in
our case is very small. Thus, we assume that the gas distribution in the
inertial reference frame is steady.

\subsection{Approach}
\label{sec:method}

We shall use the isothermal thin disk approximation in two dimensions. This
approximation is quite adequate to accretion disks in binary systems, since the
effective temperature is of the order of $10^4$~K and the characteristic ratio
of the disk thickness to its radius for this temperature is
$\lesssim 0.01$. The two-dimensional disk flow in the $(r, \phi)$ plane is
obtained by integrating the complete system of gas-dynamic equations along the
vertical coordinate $z$. The parameters of this system include the surface
density $\sigma = \int dz\,\rho$, the radial and angular velocities $u$ and
$v = \Omega r$, as well as the flat pressure $p = \int dz\,P$, where $\rho$,
$P$ are the volume density and pressure, respectively. For a perfect gas with
the adiabatic index $\gamma$, the flat pressure is a power law of the surface
density with the ‘adiabatic index’
$\gamma_\mathrm{S} = 1 + 2 (\gamma - 1) / (\gamma + 1)$
\cite{Churilov1981ATsir1157....1C,Gor'kavyj1994pprc.book.....G,Fridman1999,%
Fridman2013phgald.book}. However, in the isothermal case
$\gamma_\mathrm{S} = 1$ \cite{Gor'kavyj1994pprc.book.....G,Fridman1999} and the
flat pressure will have the same form as the volume pressure,
$p = c_\mathrm{T}^2 \sigma$. The correct reduction of the system of
three-dimensional equations to two dimensions, generally speaking, gives rise
to additional terms in equations compared to the three-dimensional
form~\cite{Gor'kavyj1994pprc.book.....G,Fridman1999}. The initial
two-dimensional system reads:
\begin{gather}
  \DPfrac{\sigma}{t} + \nabla (\sigma \vecb{V}) = 0  \;,  \\
  \DPfrac{\vecb{V}}{t} + (\vecb{V} \nabla) \vecb{V} =
  - \nabla \Phi_1 - \nabla \Phi_2
  - c_\mathrm{T}^2\,\nabla \ln (\Omega_\mathrm{K} \sigma)  \;.
\end{gather}
Here $\sigma$ is the surface density;
$\vecb{V} = \vecb{e}_r u + \vecb{e}_\phi v$;
$\Phi_1 = - G M_1 / r$,
$\Phi_2 = - G M_2 / |\vecb{R}_1 - \vecb{R}_2 + \vecb{r}|$ are gravitational
potentials of the accretor and donor, respectively;
$c_\mathrm{T}$ is the sound speed;
$\vecb{R}_1$, $\vecb{R}_2$ are radius-vectors from the barycenter to the
accretor and donor, respectively;
$\Omega_\mathrm{K} = (G M_1 / r^3)^{1/2}$ is the Keplerian angular velocity;
$G$ is the gravitational constant.

As the adopted mass ratio $M_2/M_1$ is small, the binary system barycenter lies
close to the accretor. We will assume accretor’s center at the barycenter:
$R_1 \approx 0$, $R_2 \approx A$. We also denote $M \equiv M_1$,
$q \equiv M_2/M_1$. In the cylindrical coordinates we obtain:
\begin{gather}
  \label{eq:general_continuity}
  \DPfrac{\sigma}{t} + \frac{1}{r}\,\DPfrac{(r \sigma u)}{r} +
  \frac{1}{r}\,\DPfrac{(\sigma v)}{\phi} = 0  \;,  \\
  \label{eq:general_euler_radial}
  \DPfrac{u}{t} + u\,\DPfrac{u}{r} + \frac{v}{r}\,\DPfrac{u}{\phi}
  - \frac{v^2}{r} =
  - \frac{G M}{r^2}
  - \frac{q G M\,(A \cos\phi + r)}{(A^2 + r^2 + 2 A r \cos \phi)^{3/2}}
  - c_\mathrm{T}^2\,\DPfrac{\ln (\Omega_\mathrm{K} \sigma)}{r}  \;,  \\
  \label{eq:general_euler_angular}
  \DPfrac{v}{t} + u\,\DPfrac{v}{r} + \frac{v}{r}\,\DPfrac{v}{\phi} +
  \frac{u v}{r} = - \frac{c_\mathrm{T}^2}{r}\,\DPfrac{\ln \sigma}{\phi}
  + \frac{q G M A \sin\phi}{(A^2 + r^2 + 2 A r \cos \phi)^{3/2}}  \;.
\end{gather}
Here $\phi$ is the azimuthal angle in the disk, $\phi = 0$ corresponds to the
direction towards the secondary companion along the line connecting the centers
of the binary components.

Let us set small perturbations and linearize the equations. The disk model with
precessional wave (see Section~\ref{sec:prec_wave}). will be taken as the
unperturbed solution. The perturbations can be written in the form:
$\sigma \mapsto \sigma_0 (1 + \delta)$, $u \mapsto u_0 + u$,
$v \mapsto \Omega_0 r + v$, where $|\delta| \ll 1$, $|u| \ll |u_0|$ and
$|v| \ll |\Omega_0 r|$; the values $\sigma_0$, $u_0$ and $\Omega_0$ correspond
to the unperturbed solution. Linearization of the equations results in
disappearing of terms responsible for gravitational interaction with the
donor. Equations for perturbations take the form
\begin{gather}
  \label{eq:generic_disc_delta}
  \DPfrac{\delta}{t} + u_0\,\DPfrac{\delta}{r}
  + \Omega_0\,\DPfrac{\delta}{\phi}
  + \DPfrac{u}{r}
  + \left( \DPfrac{\ln\sigma_0}{r} + \frac{1}{r} \right) u
  + \frac{1}{r}\,\DPfrac{v}{\phi}
  + \frac{1}{r}\,\DPfrac{\ln\sigma_0}{\phi}\,v = 0  \;,  \\
  \label{eq:generic_disc_u}
  \DPfrac{u}{t} + u_0\,\DPfrac{u}{r} + \Omega_0\,\DPfrac{u}{\phi}
  + \DPfrac{u_0}{r}\,u
  + \left( \frac{1}{r}\,\DPfrac{u_0}{\phi} - 2\Omega_0 \right) v
  + c_\mathrm{T}^2\,\DPfrac{\delta}{r} = 0  \;,  \\
  \label{eq:generic_disc_v}
  \DPfrac{v}{t} + u_0\,\DPfrac{v}{r} + \Omega_0\,\DPfrac{v}{\phi}
  + \left( \frac{u_0}{r} + \DPfrac{\Omega_0}{\phi} \right) v
  + \frac{\varkappa_0^2}{2\Omega_0}\,u
  + \frac{c_\mathrm{T}^2}{r}\,\DPfrac{\delta}{\phi} = 0  \;.
\end{gather}
Here $\varkappa_0^2$ is the square of the epicyclic frequency
\begin{equation}
  \varkappa_0^2 = 2\Omega_0 \left( 2\Omega_0 + r\,\Dfrac{\Omega_0}{r} \right)
  \;.
\end{equation}

Let us express perturbations in terms of harmonics
$f_\alpha(t, r, \phi) \mapsto e^{-i \omega t} f_\alpha(r, \phi)$,
where $(f_\alpha) = [\delta, u, v]$, then the system of equations
(\ref{eq:generic_disc_delta}--\ref{eq:generic_disc_v}) can be represented in
the matrix form
\begin{equation}
  \label{eq:real_representation}
  \sum_{\beta = 1}^3 A_{\alpha\beta}(r, \phi) f_\beta(r, \phi) =
  i \omega f_\alpha(r, \phi)  \;.
\end{equation}

The characteristic scale of variations of the background variables along the
radius is about $0.1 A$ (see Fig.~\ref{fig:cuts_in}), and along the angular
coordinate is about $2\pi r$. Thus, at $r \gg 0.016 A$ the angular dependence
of the background variables can be neglected. Then the problem splits in two
independent one-dimensional problems, and the density and velocity along the
corresponding radial direction in the disk can be taken as unperturbed
solutions of each of them. Assuming that the perturbations are also independent
of the angle, we obtain the system of equations for radial perturbations (see
Appendix A):
\begin{equation}
  \label{eq:radial_real_representation}
  \sum_{\beta = 1}^3 A_{\alpha\beta}(r) f_\beta(r) =
  i \omega f_\alpha(r)  \;,
\end{equation}
where the matrix of the operator $A_{\alpha\beta}$ has the form
\begin{equation}
  \label{eq:operator}
  \left[
    \begin{matrix}
      \quad u_0\,\cfrac{d}{dr} \quad &
      \cfrac{d\ln\sigma_0}{dr} + \cfrac{1}{r} + \cfrac{d}{dr} &
      0  \\[1em]
      c_\mathrm{T}^2 \cfrac{d}{dr} &
      \cfrac{du_0}{dr} + u_0\,\cfrac{d}{dr} &
      - 2 \Omega_0  \\[1em]
      0 &
      \cfrac{\varkappa_0^2}{2\Omega_0} &
      \; \cfrac{u_0}{r} + u_0\,\cfrac{d}{dr} \;
    \end{matrix}
  \right]
\end{equation}
The values $i\omega$ are eigenvalues of the operator $A_{\alpha\beta}$, and its
eigenfunctions are the solution $f_\alpha$ of the system. Thus, equation
(\ref{eq:radial_real_representation}) s a Sturm-Liouville problem.

As the unperturbed solution is sufficiently smooth, as well as to avoid
differentiation with respect to radius, we shall solve equation
(\ref{eq:radial_real_representation}) by the Galerkin method in some functional
basis representation. The problem geometry suggests that the zero-order Bessel
functions of the first kind can be conveniently chosen as the new functional
basis. In the present paper the unperturbed disk state is specified by the
numerical solution on the discrete space grid, so we use the discrete Hankel
transform for the Bessel functions in the form \cite{FiskJohnson1987}:
\begin{equation}
  \hat{f}_\alpha(k_p) =
  \sum_{q=1}^N \frac{2 J_0(k_p r_q)}{\mu_{N+1}^2 J_1^2(\mu_q)}\,
  f_\alpha(r_q)  \;,
\end{equation}
\begin{equation}
  f_\alpha(r_q) =
  \sum_{p=1}^N \frac{2 J_0(k_p r_q)}{\mu_{N+1}^2 J_1^2(\mu_p)}\,
  \hat{f}_\alpha(k_p)  \;,
\end{equation}
where $\mu_q$ is the $q$-th root of the function $J_0$; $N$ is the dimension of
the spatial grid. This transform assumes that the function $f(r)$ is given at
the finite interval $0 \leqslant r \leqslant R$ and vanishes at the interval
boundary. Its values should be calculated at the points
$r_q = R \mu_q / \mu_{N+1}$. The values of the image $\hat{f}(k_p)$ should be
calculated at points $k_p = \mu_p / R$.

Equation (\ref{eq:radial_real_representation}) can be recast to the form
\begin{equation}
  \label{eq:wave_representation}
  \sum_{s=1}^N \sum_{\beta=1}^3
  \hat{A}_{\alpha\beta}(k_p, k_s) \hat{f}_\beta(k_s) =
  i \omega \hat{f}_\alpha(k_p)  \;,
\end{equation}
where
\begin{equation}
  \hat{A}_{\alpha\beta}(k_p, k_s) = \frac{4}{\mu_{N+1}^4}
  \sum_{q=1}^N \frac{J_0(k_p r_q)}{J_1^2(\mu_q)}\,
  A_{\alpha\beta}(r_q)\,\frac{J_0(k_s r_q)}{J_1^2(\mu_q)}  \;.
\end{equation}
Note that the differential operator of the form $g(r)\,d/dr$ for an arbitrary
function should be transformed as
\begin{equation}
  g(r)\,\Dfrac{}{r} \mapsto \frac{4}{\mu_{N+1}^4}
  \sum_{q=1}^N \frac{J_0(k_p r_q)}{J_1^2(\mu_q)}\,
  \frac{k_s}{2}\,\frac{J_{-1}(k_s r_q) - J_1(k_s r_q)}{J_1^2(\mu_q)}\,
  g(r_q)  \;.
\end{equation}

Finally, to reduce the system of algebraic equations
(\ref{eq:wave_representation}) to the suitable form to which known methods of
solving the eigenvalue problem can be applied, we linearly compose elements of
the vector $\hat{f}_\alpha(k_p)$ as follows:
\begin{equation}
  (\hat{f}_I) = \left[ \hat{f}_1(k_1), \hat{f}_2(k_1), \hat{f}_3(k_1),
    \hat{f}_1(k_2), \hat{f}_2(k_2), \hat{f}_3(k_2), \dots \right]  \;.
\end{equation}
In a similar way we compose the matrix $\hat{A}_{IJ}$ to finally obtain
\begin{equation}
  \label{eq:sturm-liouville}
  \sum_{J=1}^{3N} \hat{A}_{IJ} \hat{f}_J = i \omega \hat{f}_I  \;.
\end{equation}
Eigenvalues of the matrix $\hat{A}_{IJ}$ g ive the spectrum of angular
frequencies of possible solutions, and eigenvectors give the solution in the
Bessel functions representation. For the given spatial grid dimension $N$ we
have $3N$ complex eigenfrequencies and $N$ eigenvectors for each field
($\delta$, $u$ и $v$).

\subsection{Calculation}

The method described in Section~\ref{sec:method} can be used to calculate
linear perturbations on the axially symmetric backgrounds. In the frame of the
numerical model of accretion disk considered in Section~\ref{sec:prec_wave},
this assumption, strictly speaking, is invalid. However, it is possible to
state that for each radial direction the angular dependence of all variables is
small, and therefore each radial distribution can be considered axially
symmetric. In our problem this method was independently applied to each radial
cut of the disk. The parameters of the method include: the computational domain
size $0.08~A \leqslant r \leqslant R = 0.37~A$, the grid dimension
$N = 439$. Equation (\ref{eq:sturm-liouville}) was solved using the
\textsc{lapack} library~\cite{Anderson1999LAUG}.

\begin{figure}
  \centering
  \includegraphics{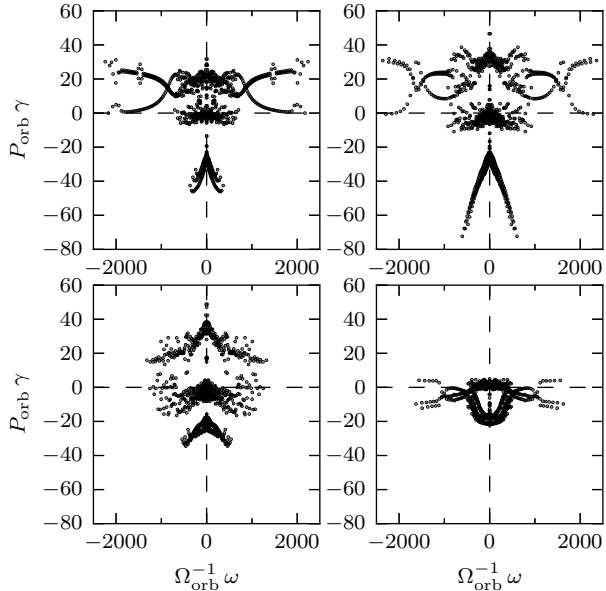}
  \caption{Maps of complex frequencies of eigensolutions in the disk cuts from
    Fig.~\ref{fig:cuts_in}.}
  \label{fig:eigenfreqs}
\end{figure}

The maps of complex frequencies for all perturbation modes are presented in
Fig.~\ref{fig:eigenfreqs}.
The maximum absolute values of angular frequencies in calculations reach
$1500~\Omega_\mathrm{orb}$, and therefore the increments lie within the range
from $-70~P_\mathrm{orb}$ to $50~P_\mathrm{orb}$. The minimal perturbation
length for the given grid dimension $N$ is around
$\lambda_N \equiv 2 R / N \approx 3 \cdot 10^{-3}~A$. This wavelength can be
compared to the Keplerian angular frequency
$(G M / \lambda_N^3)^{1/2} \approx 6 \cdot 10^3~\Omega_\mathrm{orb}$, whereas
the corresponding sound frequency is
$2\pi c_\mathrm{T} / \lambda_N \approx 40~\Omega_\mathrm{orb}$. Note that these
estimates depend on the spatial discretization scale. This scale can be related
to the maximum and minimum angular frequencies obtained in calculations.

\begin{figure}
  \centering
  \includegraphics{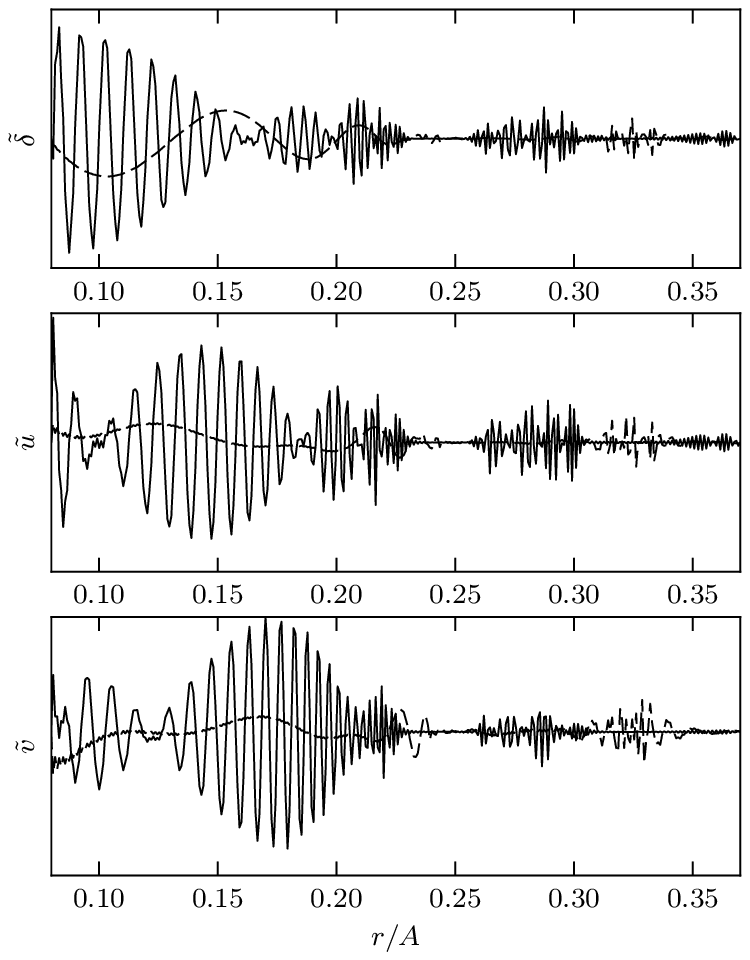}
  \includegraphics{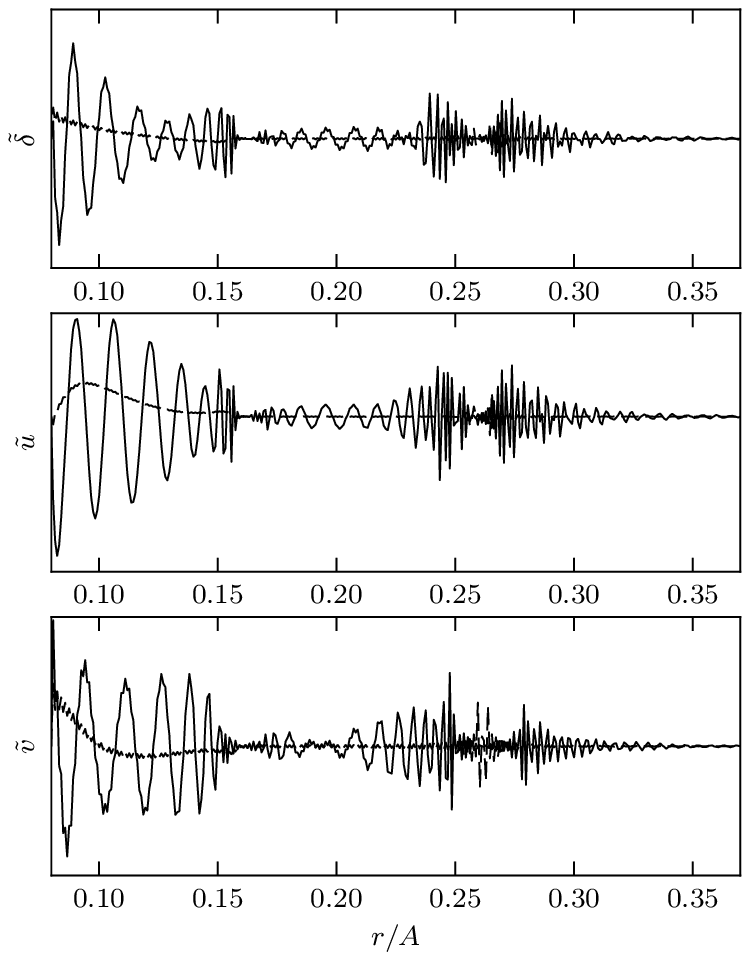}
  \caption{Real parts of the complex solutions for density, radial and angular
    velocity perturbations for some modes in the cuts (see
    Fig.~\ref{fig:cuts_in}) corresponding to $180^\circ$ (left panels) and
    $270^\circ$ (right panels). Left plots correspond to
    $\omega = 379.6~\Omega_\mathrm{orb}$, 
    $\gamma = 22.3/P_\mathrm{orb}$ (the solid curves) and
    $\omega = 35.2~\Omega_\mathrm{orb}$, $\gamma = 22.6/P_\mathrm{orb}$
    (the dashed curves). Right plots to $\omega = 158.0~\Omega_\mathrm{orb}$,
    $\gamma = 25.7/P_\mathrm{orb}$ (the solid curves) and
    $\omega = 10.4~\Omega_\mathrm{orb}$, $\gamma = 48.8/P_\mathrm{orb}$
    (the dashed curves).}
  \label{fig:eigenmodes}
\end{figure}

As the original unperturbed solution is inhomogenous, the perturbation
amplitude is different at different points of the disk. The rate of growth or
decay of perturbations is determined both by the imaginary part of the
eigenfrequency and by its local amplitude. Fig.~\ref{fig:eigenmodes}
shows real parts of eigenvectors for some modes in two cuts of the disk
corresponding to $180^\circ$ and $270^\circ$ (imaginary parts of the solutions
are different from real parts only by spatial phase).

Eigenvectors can have inhomogeneous spectral composition – the local
wavelength, being defined as the distance between maxima in different parts of
the disk, can differ by many times (see Fig.~\ref{fig:eigenfreqs}). When
examining the spectral composition of the solutions in different parts of the
disk, the wavelet-analysis may be helpful. For each mode, let us calculate the
deconvolution%
\footnote{%
In the cylindrical coordinates, the Morlet wavelet and this type of transform
are, strictly speaking, inapplicable
\cite{Kurbatov2013math.stackexchange.com/questions/373453}, but for approximate
accuracy estimates this transform turns out to be sufficient.
}
\begin{equation}
  w(r, \lambda) =
  \int dr' \left| W(r - r', \lambda)\,\delta(r') \right|  \;,
\end{equation}
where $W$ is the Morlet wavelet~\cite{Astaf'eva1996UFN} of order $5$,
\begin{equation}
  W(r, \lambda) =
  \operatorname{exp}\left[ - \frac{1}{2}%
    \left( \frac{2\pi}{5}\,\frac{r}{\lambda} - 2\pi \right)^2 \right]
  \operatorname{exp}\left( i 2\pi \frac{r}{\lambda} \right)  \;.
\end{equation}
The distribution of $w(r, \lambda)$ over wavelengths shows the characteristic
scale of change of $\lambda$ dominating around given $r$.

The turbulence viscosity coefficient $\nu_\mathrm{turb}$ or related to it the
Shakura-Sunyaev parameter $\alpha = \nu_\mathrm{turb} / (c_\mathrm{T} h)$,
where $h = c_\mathrm{T}/\Omega_\mathrm{K}$ is the semi-thickness of the
Keplerian diskis the semi-thickness of the Keplerian
disk!\cite{Shakura1973A&A....24..337S}, is the commonly accepted characteristic
of the efficiency of the angular momentum transfer in accretion disks. Although
$\nu_\mathrm{turb}$ is pertinent to well developed turbulence, in
paper~\cite{Canuto1984ApJ...280L..55C} its relation to the characteristics of
unstable liner perturbations was obtained:
\begin{equation}
  \label{eq:nu_turb}
  \nu_\mathrm{turb} = \frac{\gamma \lambda^2}{4\pi^2}  \;,
\end{equation}
where $\lambda$ and $\gamma$ are the maximum wavelength and maximum increment
for all growing modes for a given type (given branch) of perturbations. A
similar approach to the turbulent viscosity estimate was proposed in
paper~\cite{Fridman2008UFN} based on plasma theory
results~\cite[с.~299]{Kadomtsev1964vopr.teor.plazm.4}, where the instability
was considered as a consequence of the background inhomogeneity only.

\begin{figure}
  \centering
  \includegraphics{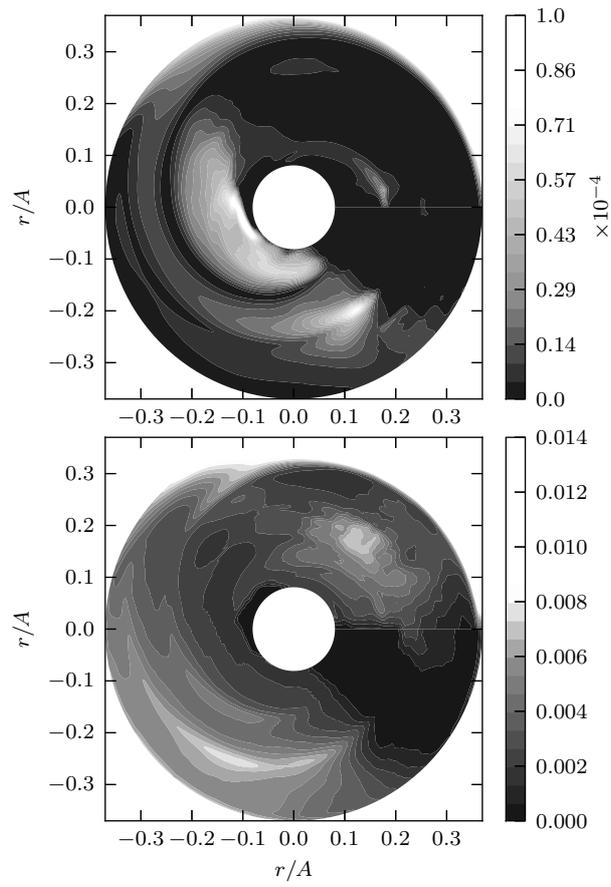}
  \caption{Perturbation amplitudes (upper panel) and the Shakura-Sunyaev
    parameter $\alpha$ (bottom panel) in the disk.}
  \label{fig:disk_out}
\end{figure}

On the eigenfrequency maps (see Fig.~\ref{fig:eigenfreqs}) it is difficult to
uniquely find perturbation branches. The estimate of $\nu_\mathrm{turb}$ using
maximum increments for all modes appears to be senseless because the modes with
maximum increments and frequencies may reflect the spatial discretization and
boundary effects. Therefore, the values of the coefficient $\nu_\mathrm{turb}$
defined by formula (\ref{eq:nu_turb}) should characterize not the whole set of
eigenvectors, but an individual mode at each point of the disk. This approach
corresponds to general equation (7) from
paper~\cite{Canuto1984ApJ...280L..55C}. In our setting, for the mode with
increment $\gamma$ and the local wavelength $\lambda$ we have
\begin{equation}
  \label{eq:alpha_old}
  \alpha = 0.21\,P_\mathrm{orb}\,\gamma\,\frac{(\lambda/A)^2}{h/A}  \;.
\end{equation}

o describe a more or less realistic accretion disk, one should specify the
power density of the family of modes and to calculate the Shalura-Sunyaev
parameter as an average over the mode ensemble. For example, in
paper~\cite{Balbus1996ApJ...467...76B} the initial conditions for the evolution
of plane waves in the outer part of an accretion disk were chosen as a
power-law power density spectrum (quadratic in the wave number) exponentially
decaying at short wavelengths. In the present problem, the plane-wavelength
approximation is invalid because the wavelength is the local variable for each
mode. The statistical weight in the mode ensemble can be specified from the
following considerations. Expression (\ref{eq:nu_turb}) is applicable to
accretion disks with taking into account two physical conditions: (i) only
three-dimensional turbulence can be induced; (ii) the characteristic growth
time of a perturbation cannot exceed one disk revolution period. These
conditions suggest two local restrictions –-- on the wavelength and increment
of a perturbation:
\begin{equation}
  \label{eq:turbulazation_conditions}
  \lambda \leqslant h  \;,\quad
  \gamma \geqslant \frac{\Omega_0}{2\pi}  \;.
\end{equation}
The final expression for the Shalura-Sunyaev parameter can be written in the
form
\begin{equation}
  \alpha_k(r) =
  \frac{\int' d\lambda\,w_k(r, \lambda)\,\alpha_k(r, \lambda)}%
       {\int' d\lambda\,w_k(r, \lambda)}  \;,
\end{equation}
for the mode with number $k$ at the point $r$ and
\begin{equation}
  \alpha(r) = \frac{{\sum_k}' w_k(r)\,\alpha_k(r)}{{\sum_k}' w_k(r)}  \;,
\end{equation}
for all modes at the point $r$. The local amplitude of perturbations for all
modes reads:
\begin{equation}
  w(r) = {\sum_k}' \int' d\lambda\,w_k(r, \lambda)  \;.
\end{equation}
Primes over the integral and sum mean that the summation is performed over the
modes and wavelengths for which conditions
(\ref{eq:turbulazation_conditions}) are fulfilled. The results of calculations
are presented in Fig.~\ref{fig:disk_out}.

\section{Physical analysis of the results}
\label{sec:physical_analysis}

The comparison of the unperturbed distributions (see Fig.~\ref{fig:disk_in} and
\ref{fig:cuts_in}) with perturbation profiles (see Fig.~\ref{fig:eigenfreqs})
eveals several features. At first, near zeros of the function $u_0$ the local
wave of perturbations decreases. At second, immediately near these points the
local amplitude of perturbations tends to zero. The first effect can be easily
explained by the following considerations. Let us maximum simplify the problem
and consider the advection part of linearized equation
(\ref{eq:generic_disc_delta}):
\begin{equation}
  \DPfrac{\delta}{t} + u_0\,\DPfrac{\delta}{r} = 0  \;.
\end{equation}
The direct substitution shows that near the root $r_\ast$ the function $u_0$
will have the form
\begin{equation}
  \delta =
  \operatorname{exp}\left( i\omega \int \frac{dr}{u_0} - i\omega t \right)
  \;.
\end{equation}
By setting the velocity variation law as $u_0 \propto r - r_\ast$, we obtain
that the local perturbation wavelength (defined as $\lambda$ in the expression
$\omega \int_r^{r+\lambda} dr/u_0 = 2\pi$) in the vicinity of $r_\ast$ behaves
as $\mathcal{O}(r - r_\ast)$.

In the immediate vicinity of zero of the function $u_0$ in equation
(\ref{eq:generic_disc_delta}) the divergence term becomes important. Let us
write equations (\ref{eq:generic_disc_delta}) and (\ref{eq:generic_disc_u}) by
ignoring geometrical terms and the tangential velocity:
\begin{gather}
  \DPfrac{\delta}{t} + u_0\,\DPfrac{\delta}{r} + \DPfrac{u}{r} = 0  \;,  \\
  \DPfrac{u}{t} + u_0\,\DPfrac{u}{r} + \DPfrac{u_0}{r}\,u
  + c_\mathrm{T}^2\,\DPfrac{\delta}{r} = 0  \;.
\end{gather}
For the background velocity variation law $u_0 \propto r - r_\ast$, in the
limit $u_0 \ll c_\mathrm{T}$ it is easy to show (by differentiating the Euler
equation with respect to radius and by excluding velocity perturbations) that
solutions take the form
$\delta \propto e^{-i\omega t} (r - r_\ast)$ and
$u \propto e^{-i\omega t} (r - r_\ast)^2$ such that in the immediate vicinity
of the point $r_\ast$ the perturbation amplitude decreases and the spatial
oscillations become ‘frozen’.

Fig.~\ref{fig:disk_out} demonstrates that the perturbation amplitude reaches
maximum in the inner part of the disk restricted by the precessional wave,
while the turbulence (assumed as large values of $\alpha$) is strongest in the
outer part of the disk. This is most clearly seen in the second and third
quadrants of the disk, where the location of the amplitude maximum coincides
with the radial velocity maximum (see Fig.~\ref{fig:disk_in}). The perturbation
amplitude decreases with decreasing velocity. In the region restricted by the
zero radial velocity, some amplitude growth is observed, especially close to
the outer edge of the precessional wave. In the fourth quadrant, where the
radial velocity in the inner part of the disk is negative (see also
Fig.~\ref{fig:cuts_in}), perturbations and turbulence are almost absent, but
there is an amplitude peak at the outer edge of the wave.

This behavior of the amplitude can be explained as follows. Let us utilize
equations (\ref{eq:generic_disc_delta}--\ref{eq:generic_disc_v}) in the
simulation box approximation whose sizes are much smaller than the distance
from the box to the center, like in paper~\cite{Balbus1996ApJ...467...76B}. Let
us represent perturbations in the form $e^{-i\omega t + ikr}$ and assume that
the perturbation length is sufficiently small to consider all unperturbed
variables as constants, as well as to neglect geometrical terms. We obtain the
equations in the form
\begin{gather}
  (-i \omega + i k u_0)\,\delta
  + \left( \DPfrac{\ln \sigma_0}{r} + i k \right) u = 0  \;,  \\
  i c_\mathrm{T}^2 k \delta
  + \left( -i \omega + i k u_0 + \DPfrac{u_0}{r} \right) u
  - 2\Omega_0 v = 0  \;,  \\
  (2 - q)\,\Omega_0 u
  + (-i \omega + i k u_0)\,v = 0  \;.
\end{gather}
The epicyclic term in the angular momentum conservation law is written as
$\varkappa_0^2/(2\Omega_0) \approx (2 - q)\,\Omega_0$, where
$q \equiv -\partial \ln \Omega_0/\partial \ln r$
\cite{Balbus1996ApJ...467...76B}. The dispersion relation for this system
reads:
\begin{equation}
  \label{eq:generic_dispersion_relation}
  \omega_\pm = u_0 k - \frac{i}{2}\,\DPfrac{u_0}{r}
  \pm \left[ c_\mathrm{T}^2 k^2
    - \frac{1}{4} \left( \DPfrac{u_0}{r} \right)^2
    + 2 (2 - q)\,\Omega_0^2
    - i c_\mathrm{T}^2\,\DPfrac{\ln \sigma_0}{r}\,k \right]^{1/2}  \;.
\end{equation}
The second term on the r.h.s. of (\ref{eq:generic_dispersion_relation}) is the
divergence term that stabilizes or destabilizes perturbations depending on
whether the flow is divergent ($\partial u_0/\partial r > 0$) as in the outer
part of the disk, or convergent ($\partial u_0/\partial r < 0$) as in the inner
part of the disk. The second term under the square root helps destabilizing the
flow –-- if the radial velocity gradient is strong enough, the perturbation
phases over the one-wavelength interval outrace each other. The third term
under the square root in the case of zero gradients would correspond to the
Rayleigh stability criterion~\cite{Balbus1996ApJ...467...76B} –-- a flow in
which the logarithmic slope of the angular velocity profile $q > 2$ is
unstable. Finally, the last term contributes to the instability increment at
all wave numbers.

Let us evaluate each term in the dispersion equation
(\ref{eq:generic_dispersion_relation}) using conditions
(\ref{eq:turbulazation_conditions}) and Fig.~\ref{fig:cuts_in}. The minimum
local wave number at a given radius must be determined by the disk
semi-thickness:
$|Ak_\mathrm{min}|^2 \equiv (2\pi A \Omega_\mathrm{K} / c_\mathrm{T})^2
\approx 10^5\,(A/r)^3$,
here everywhere but in the vicinity of roots of the radial velocity,
$|c_\mathrm{T} k_\mathrm{min}| \ll |u_0 k_\mathrm{min}|$. The term with the
velocity gradient turns out of the same order:
$|\partial(u_0/c_\mathrm{T})/\partial(r/A)|^2 \lesssim 10^5$. The term
corresponding to the Rayleigh criterion can be estimated as follows. The slope
of the rotational curve can be assumed Kepelerian everywhere (see
Fig.~\ref{fig:cuts_in}), $q = 3/2$, then
$2 (2 - q) (A \Omega_0 / c_\mathrm{T})^2 \approx 10^3\,(A/r)^3$. It is seen
that this term introduces only small stabilizing effect. The density gradient
can be estimated as
$|A^2 k_\mathrm{min}\,\partial \ln \sigma_0/\partial r| \approx %
5 \cdot 10^3\,(A/r)^{3/2}$.
Thus, the instability, at least in the simulation box approximation, can be
only due to the radial velocity gradient. The rotational motion of the gas has
only slightly stabilizing effect. The dispersion equation
(\ref{eq:generic_dispersion_relation}) as a result should have the form
\begin{equation}
  \label{eq:dispersion_relation}
  \omega_\pm = u_0 k
  \pm \left[ c_\mathrm{T}^2 k^2 - \frac{1}{4} \left( \DPfrac{u_0}{r} \right)^2
    + 2 (2 - q)\,\Omega_0^2
    \right]^{1/2}  \;.
\end{equation}

Above estimates suggest that the disk is in the boundary state between the
stability and instability. To clarify this point, we use the general integral
method of the stability analysis described in
paper~\cite{Lynden-Bell1967MNRAS.136..293L}. In this method, gas-dynamic
equations are formed through the displacement vector of gas element and the
analysis assumes this displacement to be small. In this approach, the necessary
and sufficient condition of instability in our problem can be presented as (see
Appendix B)
\begin{equation}
  \label{eq:instability_criterion}
  \left( \int dE \right)^{-1}
  \left[ \int \frac{dE}{\lambda}\,\frac{u_0}{c_\mathrm{T}} \right]^2
  + \int \frac{dE}{\lambda^2}
  \left( 1 - \frac{u_0^2}{c_\mathrm{T}^2} + \frac{\lambda^2}{h^2} \right) < 0
  \;,
\end{equation}
where $dE = dr\,r \sigma_0 |\xi|^2$; $\xi$ is the radial gas displacement;
$\lambda$ is the characteristic space scale of perturbation change. For
inequality (\ref{eq:instability_criterion}) to be satisfied it is necessary
that the first term, giving positive contribution, be small enough and the
second term be negative. The last condition is met if the scale of change of
perturbations is smaller than the disk thickness (this is one of the adopted by
us conditions of developed turbulence) and if the radial velocity in a
sufficiently large region of the disk is supersonic. The first condition is met
if the radial motion in the disk is absent or the mean value of the radial
velocity over the disk is close to zero, i.e. the velocity changes the sign. In
other terms, the velocity should take higher values and should have a large
gradient over sufficiently extended part of the disk. Both these conditions are
satisfied in the present problem (see Appendix B).

The behavior of modes near the zero radial velocity points and the dispersion
equation (\ref{eq:dispersion_relation}) are sufficient in principle to explain
many properties of unstable perturbations presented in
Fig.~\ref{fig:disk_out}. If in the inner part of the disk, restricted by the
precessional wave, the radial velocity experiences a large gradient, local
unstable modes arise. In the fourth quadrant of the disk the velocity gradient
is not too high, and additionally the alternating in sign character of the
velocity is less pronounced, so that the unstable modes are suppressed. The
sub-Keplerian gas rotation in this region (see Fig.~\ref{fig:cuts_in}) also has
the stabilizing effect.  At the precessional wave boundary, as shown above, the
perturbation amplitude vanishes. The amplitude increase in the outer parts of
the disk, in the first and forth quadrants, can be due to the boundary effects%
\footnote{%
Perturbations vanish at the calculation domain, but at the same time the
eigenvectors of the problem (\ref{eq:sturm-liouville}) have a fixed norm,
therefore in narrow regions, where there are favorable conditions for the
instability development, the perturbation amplitude can be relatively high.
}.

The physical sense of this instability can be explained as follows. In the
radial Euler equation, the term with the background velocity gradient acts on
the gas element as an external force, in addition to the pressure gradient
force and centrifugal force. In our setting, the rotational flow has
stabilizing effect and together with pressure prevents the instability
development (both corresponding terms give positive contribution under the
square root in (\ref{eq:dispersion_relation})). However, if the velocity
gradient is sufficiently high, the momentum flux transmitted to a perturbation
due to this term can exceed the counter-acting contribution from stabilizing
terms. According to (\ref{eq:dispersion_relation}), this condition occurs when
the background radial velocity change on the scale of the perturbation length
is approximately equal to or greater than the sound speed. In this case it is
possible to argue that the rear phase of the perturbation catches up with its
front phase in one wave period. Here the amplitude of perturbation maxima
increases by the mass conservation. In terms of the dispersion relation, this
signals the arising of non-zero values of the perturbation growth increment.

It should be noted once again that in our method of the instability analysis we
assume axially symmetric perturbations, while the background density and
velocity distributions do not have axial symmetry and weakly depend on the
angle with the characteristic angular scale $2\pi$ (see
Fig.~\ref{fig:disk_in}). In a real disk, perturbations with this or smaller
angular scale will shift in the tangential direction due to the background
rotation. This could weaken the perturbation growth, which is due to the radial
velocity gradient. However, the adopted here necessary condition for the
turbulence to appear (\ref{eq:turbulazation_conditions}) requires that the
characteristic growth time of perturbations be shorter than the Keplerian
time. Thus, a perturbation in its growth time cannot leave the region of the
precessional wave in the tangential direction and, hence, we observe the
instability growth.

In Section~\ref{sec:analysis}, we applied the necessary conditions of
three-dimensional turbulence development (\ref{eq:turbulazation_conditions}) to
unstable modes. In the inner part of the disk, where the angular frequency of
the gas is high, the conditions for the turbulence development are more
stringent. They become favorable only in the outer part of the disk and near
the zero radial velocity line, where the local wavelength of perturbations
decreases.

Thus, the turbulence arising in the disk can be described as
follows. Perturbations that have time to grow up to non-linear stage in less
than one disk revolution and whose wavelength does not exceed the disk
semi-thickness serve as sources for three-dimensional turbulence. The
turbulence arises predominantly along the precessional wave boundary and beyond
its outer edge and then is brought by the gas rotation and accretion flow over
the entire disk.

\section{Conclusion}

Thin Keplerian disks are known to be stable against fluid perturbations
\cite{Godon1999ApJ...521..319G,Balbus1996ApJ...467...76B}. For a hydrodynamic
instability to arise, it is necessary that the density and velocity
distributions in the disk differ from the Keplerian
ones~\cite{Balbus1996ApJ...467...76B}. The accretion disk in axially asymmetric
gravitational field of a binary stellar system provides an obvious example of
such a configuration. In paper~\cite{Bisikalo2004ARep...48..449B} we have shown
that the gravitational field of the secondary binary component excites the
precessional wave in the disk. The wave significantly changes the flow in the
disk causing the appearance of regions with large density and velocity
gradients. In the presented solution, the radial velocity gradient can be as
high as $40$ Machs.

In the present paper, we performed linear analysis of perturbations to study
the stability of an isothermal accretion disk with precessional wave. The
numerical model of the accretion disk in a binary system obtained
in~\cite{Bisikalo2004ARep...48..449B} was taken as the unperturbed background
solution. The problem of linear perturbations growth was formulated in two
dimensions in the inertial frame, where the precessional wave can be considered
as stationary. The gas flow perturbations due to time-dependent gravitational
field of the secondary component can be considered small. The strong twisting
of the precessional wave allowed us to make linear analysis of radial
perturbations only for each radial cut of the disk.

It was shown that the presence of the precessional wave gives rise to unstable
radial modes with increments up to $\sim 50/P_\mathrm{orb}$. However, for
turbulence development, the presence of background regions with large radial
velocity gradients is essential. The instability arises if the radial velocity
change in the unperturbed flow on the wavelength of a perturbation is of the
order of or greater than the sound speed. The physical reason of the
instability is that the rear phase of the perturbation starts catching up with
the frontal phase, which causes, by the mass conservation, the growth of the
perturbation maxima amplitudes. The necessary conditions for turbulence (the
wavelength smaller that the disk thickness, the increment exceeds the rotation
frequency) are fulfilled only at the precessional wave boundaries and in the
outer part of the disk.

The results of the present analysis suggest that the precessional wave in the
disk can lead to turbulence with the characteristic Shakura-Sunyaev
$\alpha$-parameter of about $0.01$.

The authors thank Yu.M. Torgashin and E.V. Polyachenko for valuable notes. We
also thank the referee, P.B. Ivanov, for important notes and comments that
helped improving the paper.

The work was supported by the Program of fundamental researches of the
Presidium of RAS P-21 and P-22, by Russian Foundation for Basic Researches
(projects 12-02-00047, 12-07-00528a, 13-02-00077) and the Program of support of
leading scientific schools of RF.

\section{Appendix A}

The method of solving linearized equations we used in the present paper ignores
the angular dependence of the background variables. This approach is different
from the ordinarily used spectral method (see, for example,
\cite{Kato2008PASJ...60..111K}), and its formulation is not fully rigorous. To
show this, expand the system (\ref{eq:real_representation}) in the basis of
some functions of the angular coordinate. Conditionally determine the
functional transformations in the form:
\begin{equation}
  \label{eq:Phi-transform}
  \tilde{f}(m) = \sum_\phi \Phi(\phi, m)\,f(\phi)  \;,
\end{equation}
\begin{equation}
  \label{eq:Psi-transform}
  f(\phi) = \sum_m \Psi(\phi, m)\,\tilde{f}(m)  \;,
\end{equation}
where $\{\Phi(\phi, m)\}$ and $\{\Psi(\phi, m)\}$ are mutually dual sets of
functions depending on the angle $\phi$ and the parameter $m$. Thus expanded
system (\ref{eq:real_representation}) takes the form
\begin{equation}
  \sum_{\beta = 1}^3 \sum_{\phi', m'}
  \Phi(\phi', m)\,A_{\alpha\beta}(r, \phi')\,\Psi(\phi', m')\,
  \tilde{f}_\beta(r, m') =
  i \omega \sum_{\phi', m'} \Phi(\phi', m)\,\Psi(\phi', m')\,
  \tilde{f}_\alpha(r, m')  \;.
\end{equation}
In the ‘standard’ approach, the expansion is doing in the orthogonal set of
functions, i.e.
$\int d\phi'\,\Phi(\phi', m)\,\Psi(\phi', m') = 0$ for
$m \neq m'$. In particular, this is valid for
$\Phi(\phi', m) \propto e^{-im\phi'}$ and
$\Psi(\phi', m') \propto e^{im'\phi'}$. The approach proposed in the present
paper relies on an expansion in two harmonics only, which we denote as $m
\equiv M$ and $m' \equiv 0$; here the orthogonality of functions is not
assumed. Expansion (\ref{eq:radial_real_representation}) can be obtained if we
define
\begin{equation}
  \label{eq:Phi}
  \Phi(\phi', M) = \delta_\mathrm{D}(\phi'-\phi)  \;,
\end{equation}
\begin{equation}
  \label{eq:Psi}
  \Psi(\phi', 0) \equiv 1  \;.
\end{equation}
where $\delta_\mathrm{D}$ is the Dirac delta-function. In this case equations
(\ref{eq:Phi-transform}) and (\ref{eq:Psi-transform}) take the form
$\tilde{f}(0) = f(\phi)$.

It is seen that $\Psi(\phi', 0)$ epresents an axially symmetric harmonic and
the operator $\int d\phi'\,\Phi(\phi', M)$ ‘cuts’ the given direction in the
disk. One of the assumptions of this approach is as follows. Although the
expansion of the system (\ref{eq:real_representation}) in non-orthogonal set of
functions is quite admissible, one cannot be sure that functions (\ref{eq:Phi})
and (\ref{eq:Psi}) belong to two mutually dual sets in the sense of definitions
(\ref{eq:Phi-transform}) and (\ref{eq:Psi-transform}) for \textit{all} $\phi$
from $0$ to $2\pi$. Despite this fact, we prefer this method since it offers a
more clear physical interpretation of the perturbation growth as a function of
the background variables distribution because it provides the angular
coordinate locality. Another assumption is that the proposed approach assumes
the use of axially symmetric modes, whereas the background solution do depends
on the angle. The distributions shown in Fig.~\ref{fig:disk_in} demonstrate
that the main angular scale of changes is $2\pi$. Therefore, this scale should
mainly contribute in perturbations, and in the standard approach this scale
would correspond to the azimuthal number $m = 1$. However, for simplicity we
assume that the perturbation is axially symmetric. Thus, the problem can be
formulated irrespective of each radial direction in the disk.

\section{Appendix B}

Below we briefly describe the instability analysis in the disk following to the
Lynden-Bell-Ostriker’s method. A detailed presentation of the method and
examples for axially symmetric flows can be found in
paper~\cite{Lynden-Bell1967MNRAS.136..293L}.

The gas-dynamic equations can be written through the displacement vector of gas
$\vecb{\xi}$ relative to its equilibrium position (marked with index ‘0’):
\begin{equation}
  \label{eq:lagrangian_shift}
  \left( \DPfrac{}{t} + \vecb{V}_0 \nabla \right)^2 \vecb{\xi} =
  - \Delta \left( c_\mathrm{T}^2 \nabla \ln \sigma + \nabla \Phi \right)  \;,
\end{equation}
where $\Delta$ is the Lagrangian difference operator determined as
\begin{align}
  \Delta f &= f(t, \vecb{r} + \vecb{\xi}(t, \vecb{r})) - f_0(t, \vecb{r})
  ={}  \\
  \label{eq:lagrangian_difference}
  &{}= f(t, \vecb{r}) - f_0(t, \vecb{r})
  + \vecb{\xi}(t, \vecb{r}) \nabla f_0(t, \vecb{r})
  + \mathcal{O}(|\vecb{\xi}|^2)  \;.
\end{align}
equation (\ref{eq:lagrangian_shift}) should be supplemented with the continuity
equation:
\begin{equation}
  \label{eq:lagrangian_shift_continuity}
  \Delta \sigma + \sigma_0 \nabla \vecb{\xi} = 0  \;.
\end{equation}
Further analysis is performed by assuming that the displacement is small and
the unperturbed disk is steady. Let us define the displacement vector as
$\vecb{\xi} \mapsto e^{i \omega t} \vecb{\xi}$. The linearized dynamical
equations have the form
\begin{equation}
  -\omega^2 \vecb{A} \vecb{\xi} + \omega \vecb{B} \vecb{\xi}
  + \vecb{C} \vecb{\xi} = 0  \;,
\end{equation}
ahere $\vecb{A}$, $\vecb{B}$, and $\vecb{C}$ are matrices composed of the
unperturbed variables and their derivatives. After multiplying this equation by
vector $\vecb{\xi}^+$ and integrating over the volume, we finally obtain the
quadratic equation for $\omega$:
\begin{equation}
  - \omega^2 \int d^2r\,\vecb{\xi}^+ \vecb{A} \vecb{\xi}
  + \omega \int d^2r\,\vecb{\xi}^+ \vecb{B} \vecb{\xi}
  + \int d^2r\,\vecb{\xi}^+ \vecb{C} \vecb{\xi} = 0
\end{equation}
or
\begin{equation}
  -\omega^2 a + \omega b + c = 0  \;.
\end{equation}
The necessary and sufficient condition for a linear perturbation to grow is the
condition for the discriminant of this equation:
\begin{equation}
  b^2 + 4 ac < 0  \;.
\end{equation}

In our problem, the displacement vector $\vecb{\xi}$ contains only the radial
component, and coefficients $a$, $b$ и $c$ have the
form~\cite{Lynden-Bell1967MNRAS.136..293L}
\begin{equation}
  a = \int dr\,r \sigma_0 |\xi|^2  \;,
\end{equation}
\begin{equation}
  b = i \int dr\,r \sigma_0
  \left( \xi^\ast\,\Dfrac{\xi}{r} - \xi\,\Dfrac{\xi^\ast}{r} \right) u_0  \;,
\end{equation}
$c = t + v + p$,
\begin{equation}
  t = \int dr\,r \sigma_0 \left( - u_0^2 \left| \Dfrac{\xi}{r} \right|^2
  + 3\Omega_0 |\xi|^2 \right)  \;,
\end{equation}
\begin{equation}
  v = \int dr\,r \sigma_0 |\xi|^2\,\Dfrac{(r \Omega_\mathrm{K}^2)}{r}  \;,
\end{equation}
\begin{equation}
  p = c_\mathrm{T}^2 \int dr\,r \sigma_0 \left| \Dfrac{\xi}{r} \right|^2  \;.
\end{equation}
Using the equilibrium flow equation
\begin{equation}
  \label{eq:equillibrium_condition}
  c_\mathrm{T}^2\,\Dfrac{\ln \sigma_0}{r}
  + r \Omega_\mathrm{K}^2 - r \Omega_0^2 = 0
\end{equation}
and the disk semi-thickness in the form $h = c_\mathrm{T} / \Omega_\mathrm{K}$,
the instability growth condition can be written as
\begin{multline}
  \label{eq:exact_instability_criterion}
  \left( \int dr\,r \sigma_0 |\xi|^2 \right)^{-1}
  \left[ \int dr\,r \sigma_0 \operatorname{Im}\!\left( \xi\,\Dfrac{\xi^\ast}{r}
    \right)
    \frac{u_0}{c_\mathrm{T}} \right]^2 +{}  \\
  {}+ \int dr\,r \sigma_0 \left| \Dfrac{\xi}{r} \right|^2
  \left( 1 - \frac{u_0^2}{c_\mathrm{T}^2} \right) +{}  \\
  {}+ \int dr\,r \sigma_0 |\xi|^2 \left( \frac{1}{h^2}
  + \frac{1}{r}\,\Dfrac{\ln \sigma_0}{r} \right) < 0  \;.
\end{multline}
Using (\ref{eq:equillibrium_condition})it is easy to show in the particular
case where the fluid is incompressible and the radial flow is absent,
expression (\ref{eq:exact_instability_criterion}) turns into the classical
Rayleigh criterion \cite{Balbus1996ApJ...467...76B}:
\begin{equation}
  \int dr\,r \sigma_0 |\xi|^2 \varkappa_0^2 < 0  \;.
\end{equation}

\begin{figure}
  \centering
  \includegraphics{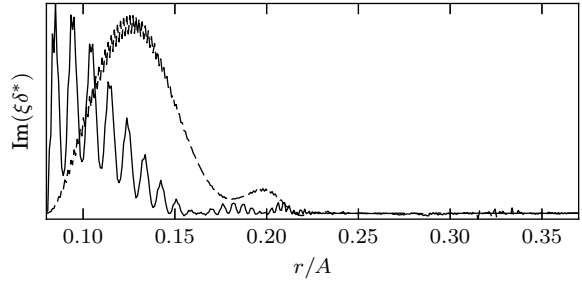}
  \caption{The value of $\operatorname{Im}(\xi \delta^\ast)$ as a function of
    the radial disk coordinate for two modes shown in the left panel of
    Fig.~\ref{fig:eigenmodes} (see Appendix B).}
  \label{fig:instability_lbo}
\end{figure}

The integrand in the first term of inequality
(\ref{eq:exact_instability_criterion}) is the product of the radial velocity
and a function whose scale of change corresponds to the given mode change
scale. Indeed, equation (\ref{eq:lagrangian_shift_continuity}) for the
displacement $\xi$ can be written as
\begin{equation}
  \Dfrac{\xi}{r} + \Dfrac{\ln (r \sigma_0)}{r}\,\xi + \delta = 0  \;,
\end{equation}
from where it follows that
$\operatorname{Im}(\xi\,d\xi^\ast/dr) = %
- \operatorname{Im}(\xi \delta^\ast)$.
The plot of this function for two modes is presented in
Fig.~\ref{fig:instability_lbo}.
By comparing these modes with those shown in Fig.~\ref{fig:eigenmodes} we
conclude that the characteristic scale of change of these functions is the
same, although functions $\operatorname{Im}(\xi\,d\xi^\ast/dr)$ apparently do
not change the sign. Oppositely, the radial velocity distribution changes the
sign. Let us define the local scale of perturbation change $\lambda$ as
$\operatorname{Im}(\xi d\xi^\ast/dr) \equiv \lambda^{-1} |\xi|^2$. As the
background distribution change scale is larger than the scale of perturbations
of interest here, the estimate $|d\xi/dr|^2 \simeq \lambda^{-2} |\xi|^2$. For
the same reason we can neglect the density gradient in the last term of the
l.h.s. of (\ref{eq:exact_instability_criterion}). Then the inequality takes the
form
\begin{equation}
  \label{eq:approx_instability_criterion}
  \left( \int dE \right)^{-1}
  \left[ \int \frac{dE}{\lambda}\,\frac{u_0}{c_\mathrm{T}} \right]^2
  + \int \frac{dE}{\lambda^2} \left( 1 - \frac{u_0^2}{c_\mathrm{T}^2}
  + \frac{\lambda^2}{h^2} \right) < 0  \;,
\end{equation}
where $dE = dr\,r \sigma_0 |\xi|^2$. The numerical check of inequality
(\ref{eq:approx_instability_criterion}) that we did also confirms the presence
of instability. For example, for modes shown in Fig.~\ref{fig:eigenmodes} the
typical value of the discriminant $1 + 4ac/b^2$ is of the order of $-1$.

\newpage

\bibliography{paper}
\bibliographystyle{unsrt}

\end{document}